\documentclass[twocolumn,graphics,prl,aps,floats,showpacs]{revtex4}
\usepackage{graphicx}
\usepackage{amsmath}
\usepackage{multirow}

\begin{document}

\title{Dispersive microwave bifurcation of a superconducting resonator
cavity incorporating a Josephson junction }
\author{E. Boaknin}
\author{V. Manucharyan}
\author{S. Fissette}
\thanks{Present address: D\'{e}partement de Physique, Universit\'{e} de
Sherbrooke, Canada}
\author{M. Metcalfe}
\author{L. Frunzio}
\author{R. Vijay}
\author{I. Siddiqi}
\thanks{Present address: Department of Physics, University of California,
Berkeley, CA, USA}
\author{A. Wallraff}
\thanks{Present address: Department of Physics, ETH Zurich, Switzerland}
\author{R. J. Schoelkopf}
\author{M. Devoret}
\affiliation{Department of Applied Physics, Yale University, New Haven, CT, USA}
\date{\today }

\begin{abstract}
We have observed the dynamical bistability of a microwave
superconducting Fabry-Perot cavity incorporating a non-linear
element in the form of Josephson tunnel junction. The effect, which
is the analog of optical bistability, manifests itself in the
transmission and reflection characteristics of the cavity and is
governed by a competition between the wave amplitude dependence of
the resonant frequency and the finite residence time of the field
energy inside the cavity. This finite residence time is solely due
to extraction of radiation from the cavity by the measurement
process. The tight quantitative agreement with a simple model based
on the Duffing oscillator equation shows that the nonlinearity, and
hence the bifurcation phenomenon, is solely dispersive.
\end{abstract}

\maketitle

Amplifying in the microwave frequency domain signals whose energy are at or
close to the quantum limit constitute an experimental challenge whose
pursuit is justified both by the fundamental understanding of amplification
mechanisms in general \cite{quantum_amplification} and by applications in
radioastronomy \cite{radioastronomy} and quantum information processing \cite%
{quantum_information}. One practical, minimal noise amplification
mechanism involves pumping a purely dispersive non-linear medium in
the vicinity of a bifurcation between two dynamical states. The weak
input signals can then be detected either in the non-hysteretic
regime for continuous, phase preserving operation or in the
hysteresis regime where the medium provides latching by switching
from one state to the other.

In this Letter, we demonstrate that a microwave superconducting
cavity incorporating a Josephson junction can display a bistable
regime of operation suitable for ultra-low noise amplification. The
bifurcation associated with this bistability involves a dispersive
non-linear dynamical evolution of the fields of the cavity on time
scales given by its quality factor rather than a slow change in its
parameters due to\ dissipation induced heating of the cavity
material. Importantly, we show that the strength of the
non-linearity depends on the combination of simple electrical
characteristics of the junction and the cavity, both being entirely
controllable by fabrication. The abscence of dissipation inside the
cavity ensures that a minimal number of modes are involved in the
amplification process, a necessary condition for amplification at
the quantum limit \cite{Caves}

\begin{figure}[tbp]
\resizebox{70mm}{!}{\includegraphics{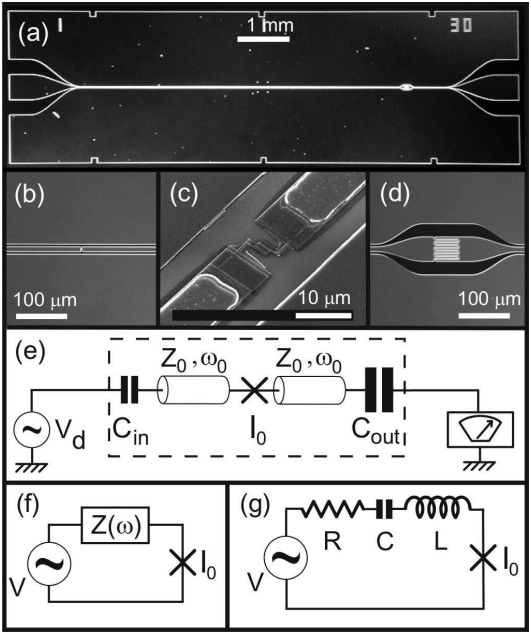}}
\caption{(a) "Dark field" photographs of a chip sample showing the
coplanar waveguide Nb resonator implementing a superconducting
Fabry-Perot cavity on chip. Asymmetric finger capacitors (b)-(d)
play the role of partially reflective mirrors. Non--dissipative
non-linearity is provided a shadow-mask evaporated Josephson
junction (c). (e) Schematic of the microwave circuit of the
experiment. The dashed box surrounds the parts which are on-chip.
Viewed from the Josephson junction the RF biasing circuit appears
(f) as an impedance $Z(\protect\omega ) $ in series with a voltage
source. (g) In the vicinity of the first mode of the resonator, the
electromagnetic environment of the junction can be modeled by an LCR
combination.} \label{picture}
\end{figure}

A Josephson tunnel junction constitutes the only radio-frequency (RF)
electrical element which is both non-dissipative and non-linear at low
temperatures and whose characteristics are engineerable in a wide parameter
range. This property makes it unique for applications in ultra-low noise
amplifiers, particle detectors, and more recently, quantum information
processing. The pioneering work on squeezing of Yurke \textit{et. al.} in
the 80's \cite{yurke1} had already showed that the Josephson junction could
achieve for microwave radiation in the quantum regime what non-linear media
could achieve for quantum optics. Since then, recent works in which a
Josephson circuit is coupled to a superconducting microwave cavity have
demonstrated that the strong-coupling regime of cavity QED in atomic physics
could be attained \cite{wallraff}. Here, we are showing that the cavity can
become so non-linear that the analog of dispersive optical bistability with
atomic ensembles \cite{kimble} can be observed.

In Fig.~\ref{picture}a we show our superconducting Fabry-Perot
cavity constructed from a coplanar waveguide transmission line by
interrupting the conductor with two in-plane capacitors playing the
role of partially transparent mirrors (see Fig.~\ref{picture}b, d).
The geometric distance between the capacitors determines the lowest
resonance frequency of this resonator. In our experiments, by
varying the length of the transmission line, the frequency is tuned
between 1.5 and 10~GHz. The quality factor $Q$ of the cavity was in
the 400-2500 range and was entirely determined by the
loading through the coupling capacitances. In a separate experiment\cite%
{Luigi} we have demonstrated that the contribution of intrinsic dissipation
to $Q$ for this type of resonator at our working temperatures ($T\sim 300$
mK) were negligible. In the middle of the cavity, the central conductor of
the transmission line is interrupted with a gap into which a Al-AlOx-Al
Josephson junction is inserted (see Fig.~\ref{picture}c), ensuring maximum
coupling between the junction and odd cavity modes. Fabrication of the
junction involves standard e-beam lithography, shadow mask evaporation and
lift-off. Good metallic contacts between the leads of the junction and the
Nb resonator were obtained with ion beam cleaning prior to the e-beam
evaporation of aluminium films. In the following, we will concentrate on the
lowest order mode for which the analysis developed in \cite{cond-mat 0612576}
applies.

\begin{figure}[tbp]
\resizebox{80mm}{!}{\includegraphics{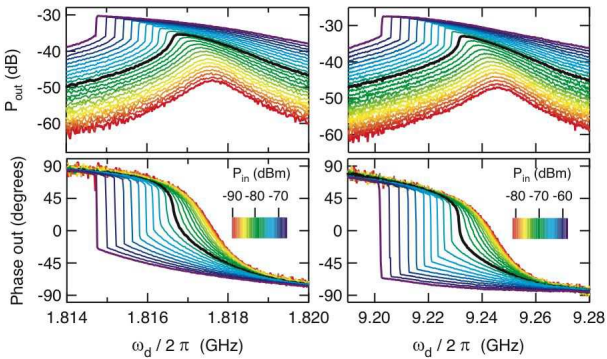}}
\caption{Measured output power (top) and phase (bottom) of transmitted
signal as a function of increasing drive frequency for a range of incoming
powers (colors) and for samples 2 (left) and 3 (right). The black line
corresponds to the critical power, separating the single state solution
regime from the bistable one. The transition between states appears as a
vertical jump.}
\label{raw-phase-amplitude}
\end{figure}
\noindent

The circuit diagram of our experiment is given in
Fig.~\ref{picture}e where the two sections of loss-less coaxial
transmission lines represent the coplanar waveguide on each side of
the junction. The transmission lines are parametrized by a
characteristic impedance ($Z_{0}\sim 50~\Omega $) and a frequency
$\omega _{0}$ corresponding to their $\lambda /2$ (fundamental)
resonance. The small input capacitor $C_{in}$ couples in a source
signal from a microwave generator while the large output capacitor
$C_{out}$ couples out the transmitted signal to the first stage of
our amplifier chain. A factor of 20 to 35 in capacitor asymmetry
ensures that all the power escaping from the resonator is collected
on the amplifier side. This maximizes the signal to noise ratio of
the experiment since it prevents the input port to participate in
the dissipation. Applying the Norton-Thevenin theorem, our microwave
circuit can be seen from the junction as an ideal RF voltage
generator connected to a series combination of the junction and a
finite linear embedding impedance $Z\left( \omega \right) $ (see
Fig.~\ref{picture}f). For drive frequencies in the vicinity of the
fundamental mode the system is very well described by the model
shown on Fig.~\ref{picture}g, where $Z\left( \omega \right) $ is now
a series LCR circuit. When combined with the junction non-linear
inductance governed by the parameter $L_{J}=\varphi _{0}/I_{0}$,
where $\varphi _{0}=\hbar /2e$, this circuit yields an
equation, analogous to the Bloch-Maxwell equations described in ref \cite%
{Mabuchi_theory}, and which can be analyzed in detail \cite{cond-mat
0612576}. In the asymmetric mirrors limit parameters of the circuit
from Fig.~\ref{picture} are given by $V=Z_{0}\omega _{0}C_{in}V_{d}$, $%
R=Z_{0}^{2}R_{L}\omega _{0}^{2}C_{out}^{2}$, $L=\frac{Z_{0}}{4~\omega
_{0}/2\pi }$, $C=\frac{1}{\pi ^{2}Z_{0}~\omega _{0}/2\pi }$ where $R_{L}$ is
the 50~$\Omega $ load resistance. Note that $\omega _{0}$ absorbs a slight
renormalization by the coupling capacitors. The quality factor is given by $%
Q=\frac{\pi }{2Z_{0}R_{L}\omega _{0}^{2}C_{2}^{2}}$. We will come back to
this model after examining the data.

\begin{figure}[tbp]
\resizebox{60mm}{!}{\includegraphics{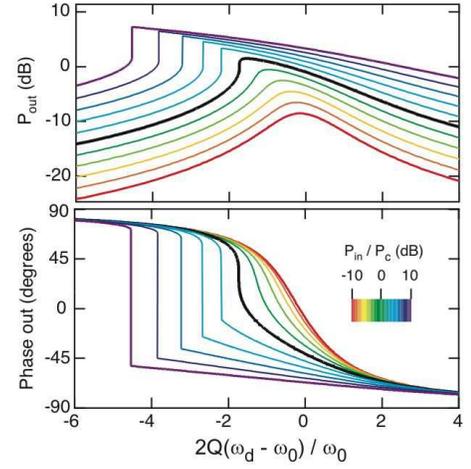}}
\caption{Theoretical predictions corresponding to the protocol of
Fig. 2 and calculated for the Duffing model, plotted as a function
of reduced drive frequency $2Q(\protect\omega _{d}-\protect\omega
_{0})/\protect\omega _{0}$.} \label{theory}
\end{figure}


The basic principle of our experiment is to measure the transmission
characteristics of the cavity as a function of both frequency and power. We
studied three different chip samples for which the relevant parameters are
given in Table~\ref{sampletable}. They were anchored to the final T=300~mK
stage of a $^{3}$He refrigerator. While we have mainly conducted CW
measurements, we have also conducted measurements in which, keeping the
frequency fixed, the power is ramped sufficiently rapidly to probe the
internal dynamics of the cavity. Since we would like to demonstrate that the
cavity is strongly non-linear even for small radiative energies stored
inside it, the challenge of the experiment is to perform precise
transmission measurements at very low power. A vector network analyzer sends
a CW microwave signal on the small capacitor side and analyzes the
transmitted signal coming out from the large capacitor side, after they have
passed through the amplifier chain which includes one cryogenic HEMT
amplifier at 4~K and two circulators placed at 300~mK and 4~K.

In Fig.~\ref{raw-phase-amplitude} we show the typical transmitted signal
amplitude (top panels) and phase (bottom panels) as a function of drive
frequency for different input drive power and for two resonator frequencies
(samples~2 and 3). The drive frequency was swept in 800~ms. At low drive
power $P_{in}=\frac{1}{2R_{L}}V_{d}^{2}$ the response of the resonator is a
Lorentzian centered on the resonator frequency. As drive power is increased,
the resonant response sharpens on the low frequency side and softens on the
high frequency side. At a critical drive power $P_{in}=P_{c}$, the maximum
slope on the low frequency side diverges. Upon increasing the drive power
further, a abrupt transition develops, whose position shifts to lower
frequency with increasing power. The two panels of Fig.~\ref%
{raw-phase-amplitude} demonstrate that the phenomenon presents itself in
identical manners for two resonators despite their different resonant
frequencies and quality factors. For clarity the response at the critical
power is highlighted in black. At even higher drive power (data not shown),
the system becomes chaotic as described elsewhere \cite{JBA1}.

\begin{table}[tbp]
\begin{center}
\begin{tabular}{c|c|c|c|c}
\hline\hline
&  & sample 1 & sample 2 & sample 3 \\ \hline\hline
\multirow{4}{*}{measured} & $\omega _{0}$(GHz) & $1.828$ & $1.8177$ & $9.246$
\\
& $Q$ & $2360$ & $1500$ & $460$ \\
& $C_{in}$(pF) & $2.3\pm 0.2$ & $1.5\pm 0.5$ & $1.0 \pm 0.1$ \\
& $P_{c}$(dBm) & $-85.5\pm 1$ & $-76.5\pm 1$ & $-70\pm 1.5$ \\ \hline
\multirow{4}{*}{inferred} & $L$ (nH) & $6.7$ & $6.8$ & $1.3$ \\
& $C$ (pF) & $1100$ & $1100$ & $200$ \\
& $R$~$(\Omega )$ & $0.034$ & $0.053$ & $0.17$ \\
& $I_{0}$ ($\mu $A) & $1.6\pm 0.2$ & $1.5\pm 0.5$ & $2.9\pm 0.35$ \\ \hline
\multirow{1}{*}{expected} & $I_{0}$ ($\mu $A) & $1.3\pm 0.1$ & $1.6\pm 0.1$
& $3.35\pm 0.15$ \\ \hline\hline
\end{tabular}%
\end{center}
\caption{Table showing the measured values of parameters for samples studied
in experiment, as well as inferred and expected parameters. $C_{in}$ is
obtained either from the measurement of other identical resonators with no
Josephson junctions (samples 1 and 2) or by measuring the ratio of input and
output powers leading to criticality and assuming that the quality factor is
set by $C_{out}$ (sample 3).}
\label{sampletable}
\end{table}

\begin{figure}[tbp]
\resizebox{80mm}{!}{\includegraphics{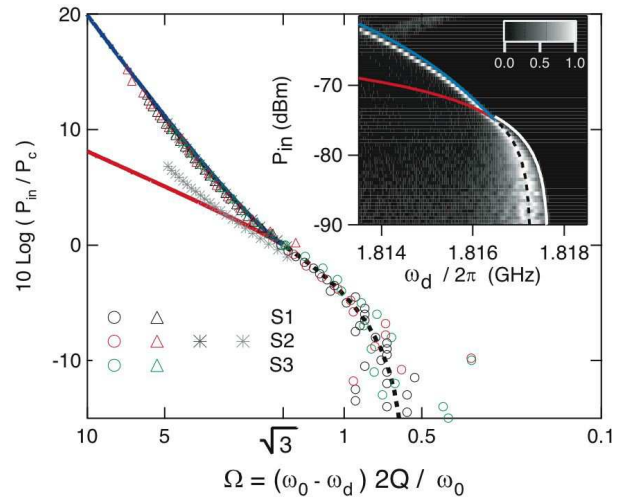}}
\caption{Logarithmic plot of the highest derivative $dP_{out}/d\protect%
\omega _{d}$ \ as a function of reduced parameters $\Omega $ and $%
P_{in}/P_{c}$. Lines depict theoretical prediction for the Duffing model.
Triangles, circles and crosses represent respectively the measured
bifurcation power, its highest derivative (below $P_{c}$), and the data of
the hysteresis measurement. \emph{Inset:} derivative of the output power
with respect to drive frequency, normalized to its maximum as a function of
drive frequency and power. Note the white line delimits the point where the
output power is maximum (the derivative changes sign).}
\label{universal}
\end{figure}
\noindent

These results should be fully governed by the static solutions of the
Duffing oscillator as described in \cite{cond-mat 0612576}. Indeed, simple
analytical relations for the transmitted power and phase at different drive
powers show excellent qualitative agreement with the measurements as shown
in Fig.~\ref{theory}. We now focus on the locations in the $\left( \omega
_{d},P_{in}\right) $ plane where the susceptibility is maximum, i.e.
locations where $\mathrm{d}P_{out}/\mathrm{d}\omega _{d}$ is maximum. When $%
P_{in}>P_{c}$, these should correspond with the bifurcation points.
We will also follow the point of maximum $P_{out}$. These results
are plotted for the three samples and three different values of the
DC critical current of the Josephson junctions patterned in our
samples. We plot them against the reduced frequency $\Omega \equiv
2Q(\omega _{0}-\omega _{d})/\omega _{0}$ and $P_{in}/P_{c}$, both
obtained independently for each sample. As expected, the various
experimental points fall on a universal theoretical curve when
properly scaled. Based on \cite{cond-mat 0612576} we expect the
bifurcation power to be given by $P_{b}(\Omega
)/P_{c}=\frac{1}{12\sqrt{3}}\Omega ^{3}(1+9/\Omega ^{2}\mp
(1-3/\Omega ^{2})^{2/3})$ (blue and red lines respectively) and the
power of highest
derivative below $P_{c}$ to be $P_{HD}/P_{c}=\frac{\sqrt{3}}{2}\Omega -\frac{%
1}{2}$ (dashed line). They are also shown on a plot of the output power in
the inset of Fig.~\ref{universal}. There, $\mathrm{d}P_{out}/\mathrm{d}%
\omega _{d}$ is normalized to its maximum and plotted as a function of the
drive frequency and power. The maximum output power (below $P_{c}$) is shown
as a white line defined by $P_{max}/P_{c}=\frac{9}{8\sqrt{3}}\Omega $ and
coincides with the change of sign of $\mathrm{d}P_{out}/\mathrm{d}\omega _{d}
$. Note that these measurements do not probe the hysteresis since the
frequency is swept only in the forward direction. To verify the hysteretic
behavior of the phenomenon, we used triangular power sweeps for several
frequencies. We were able to probe the power and frequency dependence of
both bifurcation points. The resulting data is shown as stars on Fig.~\ref%
{universal}. The deviation from predictions is most likely due to the
proximity of the lower bifurcation current amplitude to $I_{0}$, situation
not well addressed by the Duffing model. Note that the upper and lower
bifurcation current in this RF experiment correspond to the switching and
retrapping current, respectively, in DC Josephson IV measurement.

We now discuss how, from the measured value $P_{c}$ of the critical power,
we can infer $I_{0}$. In contrast to a DC experiment, where one can
characterize the biasing circuit with great precision, an absolute
calibration of the environmental impedance $Z(\omega )$ at RF frequencies is
arduous. From the values of $C_{in}$, $Z_{0}$, $L$ and $Q$ obtained from the
low power (linear) measurements, as well as the relations $V_{d~crit}=\frac{%
V_{C}}{Z_{0}C_{in}\omega _{0}}$ and $V_{c}=\frac{8}{3^{3/4}}(\frac{L+L_{J}}{%
L_{J}Q})^{3/2}\varphi _{0}\omega _{0}$, we extract $I_{0}=\varphi _{0}/L_{J}$
from the measured value of $P_{c}=V_{d~crit}^{2}/2R_{L}$. In table 1 we show
that $I_{0}$ obtained in this way is consistent with the normal resistance
test of junctions fabricated in the same batch. We can thus verify that the
Duffing model hypothesis $I_{c}/I_{0}=\frac{4}{\sqrt[4]{3}}\sqrt{\frac{%
L+L_{J}}{L_{J}}\frac{1}{Q}}\ll 1$ for the RF critical current amplitude $%
I_{c}$ is well satisfied \cite{cond-mat 0612576}.

Other experiments on an AC biased Josephson junction have been realized \cite%
{yurke1,JBA2,janicelee,lupascu}, albeit with a lower degree of
control over the electrodynamic environment. Nonlinear effects
involving superconducting weak links in resonators have been
reported~\cite{Nbweaklinkres, Buks, Haviland}. Moreover, the Duffing
oscillator physics has been shown to appear in several other systems
including nanomechanical resonators \cite{mechanical} and relativistic electrons \cite%
{Gabrielse}. However, our realization offers an unprecedented
opportunity to study dynamical driven systems in the regime where
quantum fluctuations are dominant \cite{Mabuchi_theory},
\cite{dykman}. This regime is yet unexplored experimentally.

The excellent overall agreement between experimental results and
theoretical predictions allows us to eliminate non-linear
dissipative effects as the cause for the bifurcation. For instance,
if the dissipation in the resonator would increase with power the
resonance would become broader, its maximum would decrease and the
resonance curves will eventually cross each other, an effect never
observed in our experiment.

The main application we envisage in the short term for this phenomenon is
parametric amplification at microwave frequency which would reach the
quantum limit at the practical level, not simply at the proof-of-principle
level. It could also used as a superconducting quantum bit readout \cite%
{JBAqubit}. Because the cavity frequency is mainly determined by geometry,
we can stage in the frequency domain several cavities and multiplex the
measurement or readout of a collection of entities.

In conclusion, we have observed the dynamical bifurcation of a
superconducting microwave cavity incorporating a non-linear element in the
form a Josephson tunnel junction. This bifurcation is analogous to the
optical bistability observed with atoms in a Fabry-Perot in QED experiments.
Our system is therefore at the crossroads between the physics of dynamical
systems and quantum optics. Our level of control on the various parameters
of this bistable phenomenon opens the door to amplification at the quantum
limit, squeezing and other analog quantum information processing functions
in the strong and ultra-strong coupling regimes.

The authors are grateful to Daniel Esteve, Denis Vion and Steve
Girvin for helpful discussions. This work was supported by NSA
through ARO Grant No. W911NF-05-1-0365, the Keck Foundation, and the
NSF through Grant No. DMR-0325580.

\end{document}